\documentclass[graybox]{svmult}

\usepackage{mathptmx}       % selects Times Roman as basic font
\usepackage{helvet}         % selects Helvetica as sans-serif font
\usepackage{courier}        % selects Courier as typewriter font
\usepackage{type1cm}        % activate if the above 3 fonts are
\usepackage{makeidx}         % allows index generation
\usepackage{graphicx}        % standard LaTeX graphics tool
\usepackage{multicol}        % used for the two-column index
\usepackage[bottom]{footmisc}% places footnotes at page bottom

\usepackage{graphicx,amsbsy,dcolumn}
\usepackage{amsmath,amsfonts,euscript,epsfig,color}
\usepackage{mathrsfs} % $$\mathscr{B}$$

\makeindex             % used for the subject index
\begin{document}

\title*{Tilted Lema\^itre model and the dark flow}
\author{Julio J. Fern\'andez and J.-F. Pascual-S\'anchez}
\institute{Julio J. Fern\'andez \at Dept. Fisica Te\'orica, Facultad de Ciencias,
Universidad de Valladolid, Valladolid, 47011, Spain,
\email{julio.j.fernandez@hotmail.es} \and  J.-F. Pascual-S\'anchez \at IMUVA
and Dept. Matem\'atica Aplicada, E.I.I., Universidad de
Valladolid, 47011, Spain, \email{jfpascua@maf.uva.es}}
\maketitle

\abstract*{In the last years, the peculiar velocities of many  X-ray galaxies clusters with respect to the distance have been measured directly in the rest frame of the cosmic microwave background radiation (CBR), using the kinematic Sunyaev-Zeldovich (kSZ) effect. These  measures prove that exists a highly coherent motion, extending out to at least to  $~1\, Gpc$, of the matter rest frame  with respect to the CBR rest frame. This global motion was named ``dark flow''.  By using an inhomogeneous spherically symmetric  ``tilted'' Lema\^itre model, we could explain the dark flow if we assume a linear increase with distance of the peculiar velocities, which is in principle allowed by these observations. This linear increase of the dark flow with the distance has the same behavior that the intrinsic dipole, due to the kinematic acceleration, which appears in the Hubble law of the Lema\^itre model. In the ``tilted'' Lema\^itre model considered, we consider that  the radiation orthogonal congruence  is a perfect fluid and the matter ``tilted'' congruence is an imperfect fluid  with heat flux.}

\abstract{In the last years, the peculiar velocities of many  X-ray galaxies clusters with respect to the distance have been measured directly in the rest frame of the cosmic microwave background radiation (CBR), using the kinematic Sunyaev-Zeldovich (kSZ) effect. These  measures prove that exists a highly coherent motion, extending out to at least to  $~1\, Gpc$, of the matter rest frame  with respect to the CBR rest frame. This global motion was named ``dark flow''.  By using an inhomogeneous spherically symmetric  ``tilted'' Lema\^itre model, we could explain the dark flow if we assume a linear increase with distance of the peculiar velocities, which is in principle allowed by these observations. This linear increase of the dark flow with the distance has the same behavior that the intrinsic dipole, due to the kinematic acceleration, which appears in the Hubble law of the Lema\^itre model. In the ``tilted'' Lema\^itre model considered, we consider that  the radiation orthogonal congruence  is a perfect fluid and the matter ``tilted'' congruence is an imperfect fluid  with heat flux.}

\section{Tilted cosmological models}
\label{sec:1}

In cosmology it is essential to specify the set of observers, or
rather, the congruences of world-lines from which
observations are made.

Cosmological quantities depend on the choice of these congruences, specified by 4-velocities fields.
For instance, with the same spacetime  metric, one can have  two
         different stress-energy-momentum tensors, corresponding to two different congruences,
          both satisfying the Einstein equations.
          The interpretation of the universe of two observers,
                  associated with two different congruences, can be  radically different.
                  These models are named ``tilted" in the literature, which begins with \cite{ellis1}.

If  $\Sigma$ is a global 3D spacelike  space, which exists assuming zero vorticity,
   $n{^{*}}^{\alpha}$  is the four velocity of the normal or orthogonal congruence to $\Sigma$ and $u^{\alpha}$   the four velocity of the tilted one.

The relationship bewteen both velocities is, at low speed, a galilean transformation:
        $ u^{\alpha}=n{^{*}}^{\alpha}+ v^{\alpha}$.
 The relative (``tilting'') peculiar velocity $v^{\alpha}$ between the two congruences may be related to a physical phenomenon such as the observed motion of our galaxy relative to the cosmic microwave background radiation, the CBR dipole, which is usually interpreted as a  Doppler effect.

Previous works on tilted models have been realized by using  FLRW \cite{coley1}, Bianchi \cite{ellis1} and, recently, Lema\^itre-Tolman-Bondi (LTB) \cite{herre1} and  Szekeres metrics \cite{herre2}.

 In this work and in the more detailed article \cite{nos}, we will consider a different case: the Lema\^itre model. This is the generalization of the LTB model to the case of non-dust matter with a non-null  pressure gradient, which gives rise to a kinematic acceleration. This  can explain the acceleration of the expansion obtained by the SN1a supernovae distance measures, without considering dark energy, see \cite{pas1}.

\section{The tilted Lema\^itre model}
\label{sec:2}

The metric of the  Lema\^itre model
 in comoving coordinates is:
             \begin{equation}
ds^2=-N(r,t)^2\,dt^2+B(r,t)^2\,dr^2+ R(r,t)^2\,(d\theta^2+\,sin^2\theta d\phi^2),
\end{equation}

Where $N(r,t)$ is the lapse and $R(r,t)$ is the areal radius (or warp factor).
It is spherically symmetric, with three Killing vectors. Its symmetry group is a $G_{3}/S^{2}$.
and it has a ``local''  preferred radial spatial direction at each point.
            Also, it belongs to  Petrov type D and it has null magnetic Weyl tensor.

This metric is compatible with a general non-perfect (with heat flux and anisotropic pressures)
        comoving fluid, as Lema\^itre first pointed out in \cite{lema1}.

 The  comoving congruence is normal to a foliation of global 3D spacelike hypersurfaces,
 $n{^{*}}_{\alpha}=N^{-1}\delta_{\alpha}^t $.

  The kinematical quantities, expansion, shear tensor and acceleration four-vector of the Lema\^itre spacetime are all non-null and can be reduced to scalars.
   Moreover, as the vorticity $\omega{^{*}}_{\alpha\beta} = 0$, it admits  a global 1+3 splitting. Also, as  a LRSIIb model,  it admits a 1+1+2 threading.

Choosing the radially ``tilted'' non-comoving congruence $ u^{\alpha}$ as
\begin{equation}
u^{\alpha}= \left(\frac{1}{N},\frac{v_{r}}{B },0,0\right),
\end{equation}
where $v_{r}$ means the radial peculiar velocity w.r.t. the normal $n{^{*}}^{\alpha}$ frame.
 Due the spherical  symmetry of the Lema\^itre model, $v_{r}$ is a spherical average.

All the kinematical quantities of the ``tilted'' Lema\^itre spacetime w.r.t. the ``tilted'' congruence $u^{\alpha}$
can be computed and related to the non-tilted ones.

Consider the Lema\^itre metric $g_{\alpha\beta}$ as the solution of the Einstein equations for the two different fluid congruences. Where the two stress-energy-momentum tensors $T^{*}_{\alpha \beta}$ and $T_{\alpha \beta}$ are:
\begin{equation}
 T^{*}_{\alpha\beta}=\frac{4}{3}\mu^{*} n^{*}_{\alpha}n^{*}_{\beta}+ \frac{1}{3} \mu^{*}\, g_{\alpha\beta},
 \end{equation}
which corresponds to an inhomogeneous  radiation fluid with energy density $\mu^{*}(r,t)$      and
\begin{equation}
              T_{\alpha\beta}=(\mu +p)u_{\alpha}u_{\beta}+p\,g_{\alpha\beta}+q_{\alpha}u_{\beta}+u_{\alpha}q_{\beta},
              \end{equation}
which corresponds to an inhomogeneous   imperfect matter fluid, with heat flux,
$q_{\alpha}= q\,s_{\alpha}(r,t)$, in the radial direction of the observer.

Since the Einstein tensor  is the same for the tilted (matter) and orthogonal (radiation) congruences, imposing
$$T_{\alpha \beta}= T^{*}_{\alpha \beta}\, ,$$
we should have the following relations between dynamical quantities:\\
\begin{equation}
  \mu = \mu^{*} (1+  \frac{4}{3}  v_{r}^2) \,;\,\,\,
 p = \mu^{*} (\frac{1}{3} + \frac{4}{3}  v_{r}^2) \,;\,\,\,
  q = \frac{4}{3} v_{r} .\\
\end{equation}

 \section{The CBR dipole and the dark flow}
 \label{sec:3}

 In the $\Lambda CDM$ model the Hubble expansion is  assumed to be uniform, so that the differences between  peculiar velocities of galaxies (or clusters) $v_{r}$ and the observer velocity, are   deviations from the isotropy of the usual Hubble law of the  FLRW models.
In the concordance $\Lambda CDM$ model,  the spherically averaged peculiar bulk
velocity has a hyperbolic $1/r$ dependence with distance.

 On the other hand, the kinematic Sunyaev-Zeldovich (kSZ) effect, which measures the dipole anisotropy of the CBR through a tiny temperature shift (of the order of $\mu K$) in the spectrum of CMB photons scattered from hot gas in clusters of galaxies, gives us the peculiar velocity of any cluster directly in the rest frame of the CMB.
  By using the kSZ effect, the authors in the review \cite{atrio}  claim to have detected a highly significant CBR dipole for $\sim1200$  clusters with redshift $z \leq 0.12$ up to $z = 0.6$.
\begin{figure}
    \begin{center}
      \includegraphics[width=9cm,angle=0]{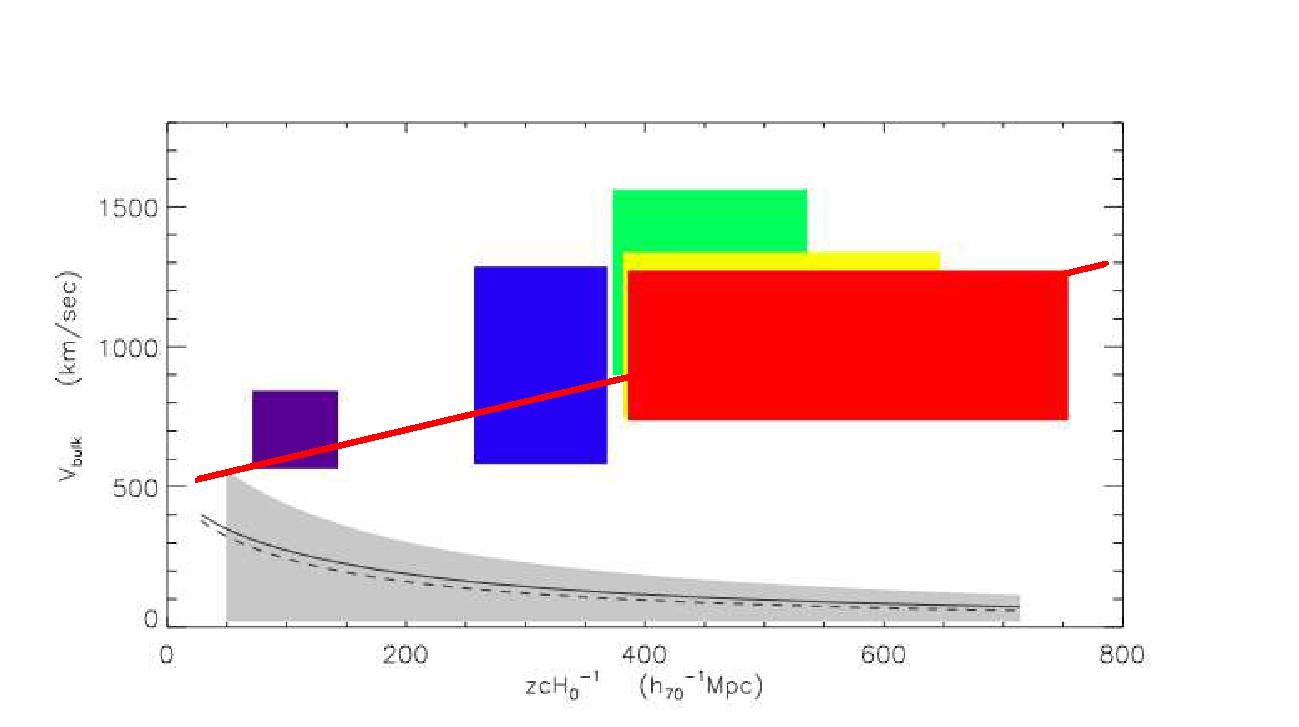}
          \caption{In  colored rectangles, the kSZ measured peculiar velocities as a function of the distance.
    Shaded grey region shows the peculiar velocity in the concordance $\Lambda CDM$ model.
    The red line is our hypotesis of linear increasing with the distance of the kinematic dipole of the dark flow.
    Figure taken, without the red line, from A. Kashlinsky et al., \cite{atrio}.}  \label{figure1}
    \end{center}
    \end{figure}
  This detection proves that exists a highly coherent motion, extending out to at least to  $~1\, Gpc$, of the matter rest frame  with respect to the CBR rest frame congruence. This global motion was named ``dark flow'' by the authors of \cite{atrio} and supposed by them to be approximately constant, but see  Fig.~\ref{figure1}. This dark flow do not has the $1/r$ dependence with distance, so this is in contradiction with the result of the  $\Lambda CDM$ model.
   Also, it appears that the dark flow may extend across the entire observable Universe horizon. This can be considered as an evidence for a ``tilted'' Universe.
   In \cite{pas} it was shown that, in the Lema\^itre model,  the following generalized Hubble law is verified:
 \begin{equation}
     c\,z = \left( \frac{\dot{R}}{R}-A\cos \Psi +\sqrt{3}\,\sigma\cos ^2 \Psi\right)_0\, d_{a}.
     \end{equation}
     This law shows an intrinsic  dipole, due to the acceleration, increasing linearly with the distance as in our hypothesis.
Where,
 $\Psi$ is the angle between the direction of observation
of a light ray and the preferred  radial vector $e_r$ of the Lema\^itre model, $\sigma$ is the
scalar shear, $A$ is the kinematic acceleration and $d_{a}$ is the angular diameter distance. However, up to now,  there is not the necessary accuracy to obtain the possible shear  using  the kSZ effect. Note that in the pure dust
 Lema\^itre-Tolman-Bondi model the intrinsic  dipole, due to the acceleration, is absent.

In conclusion, accepting that the dark flow is real and that it increases linearly with the distance,
the tilted Lema\^itre model could be considered as a candidate to explain it.
Then there  exists a preferred radial spatial direction in the Universe, given by the matter dark flow. Is this ``the axis of evil'' or better ``the rebel or guerrillero  axis'', because it reappears when the dark flow is considered?.

\end{document}